\documentclass[conference]{IEEEtran}
\IEEEoverridecommandlockouts
\usepackage{cite}
\usepackage{amsmath,amssymb,amsfonts}
\usepackage{algorithmic}
\usepackage{graphicx}
\usepackage{textcomp}
\usepackage{xcolor}
\usepackage [english]{babel}
\usepackage [autostyle, english = american]{csquotes}
\MakeOuterQuote{"}
\usepackage{graphicx}
\usepackage{float}
\usepackage{graphics} % for pdf, bitmapped graphics files
\usepackage{epsfig} 
\usepackage{subfigure}
\usepackage{tabularx}
\usepackage{multicol}
\usepackage{multirow}
\usepackage{tabularx}
\usepackage{pdflscape}
\usepackage{bbding}
\usepackage{wasysym}
\usepackage{amssymb}% http://ctan.org/pkg/amssymb
\usepackage{pifont}% http://ctan.org/pkg/pifont
\usepackage{amssymb}
\usepackage{afterpage}
\usepackage{capt-of}
\usepackage{rotating}
\usepackage{lscape}
\usepackage{hyperref}
\usepackage{amsmath}
\usepackage{pgfplots}
\pgfplotsset{width=8cm,compat=1.9}
\graphicspath{ {images/} }
\usepackage{amsmath}

\def\BibTeX{{\rm B\kern-.05em{\sc i\kern-.025em b}\kern-.08em
    T\kern-.1667em\lower.7ex\hbox{E}\kern-.125emX}}
    
\makeatletter
\patchcmd{\@maketitle}
  {\addvspace{0.5\baselineskip}\egroup}
  {\addvspace{-1\baselineskip}\egroup}
  {}
  {}
\makeatother

\begin{document}

\title{APEX: Adaptive Ext4 File System for Enhanced Data Recoverability in Edge Devices}

\author{\IEEEauthorblockN{Shreshth Tuli}
\IEEEauthorblockA{
% \textit{Department of Computer Science and Engineering} \\
\textit{Indian Institute of Technology}\\
Delhi, India\\
shreshthtuli@gmail.com
}
\and
\IEEEauthorblockN{Shikhar Tuli}
\IEEEauthorblockA{
% \textit{Department of Computer Science and Engineering} \\
\textit{Indian Institute of Technology}\\
Delhi, India\\
shikhartuli98@gmail.com
}
\and
\IEEEauthorblockN{Udit Jain}
\IEEEauthorblockA{
% \textit{Department of Computer Science and Engineering} \\
\textit{Indian Institute of Technology}\\
Delhi, India\\
uditjain.iitd@gmail.com
}

% \and
% \IEEEauthorblockN{Preeti Ranjan Panda}
% \IEEEauthorblockA{
% % \textit{Department of Computer Science and Engineering} \\
% \textit{Indian Institute of Technology}\\
% Delhi, India\\
% panda@iitd.ac.in
% }
\and
\IEEEauthorblockN{Rajkumar Buyya}
\IEEEauthorblockA{
% \textit{Department of Computer Science and Engineering} \\
\textit{The University of Melbourne}\\
Melbourne, Australia\\
rbuyya@unimelb.edu.au
}
}

% \author{\IEEEauthorblockN{Shreshth Tuli\IEEEauthorrefmark{1},
% Udit Jain\IEEEauthorrefmark{1}, Shikhar Tuli\IEEEauthorrefmark{1}, Preeti Ranjan Panda\IEEEauthorrefmark{1} and
% Rajkumar Buyya\IEEEauthorrefmark{2}}
% \IEEEauthorblockA{
% \IEEEauthorrefmark{1} Indian Institute of Technology Delhi, India,
% \IEEEauthorrefmark{2} University of Melbourne, Australia}}

\maketitle

\begin{abstract}
Recently Edge Computing paradigm has gained significant popularity both in industry and academia. With its increased usage in real-life scenarios, security, privacy and integrity of data in such environments have become critical. Malicious deletion of mission-critical data due to ransomware, trojans and viruses has been a huge menace and recovering such lost data is an active field of research. As most of Edge computing devices have compute and storage limitations, difficult constraints arise in providing an optimal scheme for data protection. These devices mostly use Linux/Unix based operating systems. Hence, this work focuses on extending the Ext4 file system to APEX (Adaptive Ext4): a file system based on novel on-the-fly learning model that provides an Adaptive Recover-ability Aware file allocation platform for efficient post-deletion data recovery and therefore maintaining data integrity. Our recovery model and its lightweight implementation allow significant improvement in recover-ability of lost data with lower compute, space, time, and cost overheads compared to other methods. We demonstrate the effectiveness of APEX through a case study of overwriting surveillance videos by CryPy malware on Raspberry-Pi based Edge deployment and show 678\% and 32\% higher recovery than Ext4 and current state-of-the-art File Systems. We also evaluate the overhead characteristics and experimentally show that they are lower than other related works.
\end{abstract}

%%%%%%%%%%%%%%%%%%%%%%%%%%%%%%%%%%%%%%%%%%%%%%%%%%%%%%%%%%%%%%%%%%%%%%%%%%%%%%%%
\section{Introduction}

Internet of Things (IoT) paradigm enables integration of different sensors, compute resources and actuators to perceive external environment and act to provide utility in different applications like healthcare, transportation, surveillance among others \cite{iot}. The original idea of providing enhanced Quality of Service (QoS), which is a measure of the performance of such services, was to provide distributed computation and computation on Cloud \cite{cloud-hist}. A new computing paradigm namely `Fog Computing' leverages resources both at public cloud and the edge of the network. It provides resource abundant machines available at multi-hop distance and resource constrained machines close to the user. Several works have shown that Fog computing can provide better and cheaper solutions compared to only Cloud based approaches \cite{fog, FogBus}. Thus, many of such IoT systems have recently been realized using Edge/Fog computing frameworks \cite{iot-demand}. Due to the increased usage of such devices and frameworks, there has been increasing interest in developing efficient techniques to provide improved QoS that such systems provide \cite{qos}. 

 Edge is considered as a nontrivial extension of the cloud, therefore, it is inevitable that same security and privacy challenges will persist \cite{security-and-privacy-fog}. Many works in literature show that there exist new security threats in Edge paradigm which provide opportunities of developing more robust and better systems \cite{fog-sec3}. One of the most crucial security problems is the loss of critical data due to malicious entities \cite{fog-sec3}. Some ways used by hackers to corrupt, delete and steal such critical data include the usage of data stealing malwares, ransomwares, trojans, deleting or overwriting viruses. Most malware attacks have been based on stealing crucial data from the system and requesting ransom payments in return for the data. Unfortunately, according to a recent survey \cite{ransom-survey}, the number of novel ransomwares/trojans has increased up to 50 times in the last decade amounting to millions of dollars of illicit revenue. 
 
 Preventing such attacks has not been popular in IoT. This is because it requires significant computation and space  \cite{ransom-exp} that drastically increases the cost of IoT deployment which is unfavorable for users. Also, it has been shown that no matter what protection mechanisms are put in place, edge paradigms will be successfully attacked \cite{edge-attacks}. Thus, the detection and recovery from such attacks seems a critical requirement for Edge Computing domain. Due to increasing frequency of novel attacks and their types, detection is quite challenging \cite{cyber-attack-chal}. Despite such challenges, there exist prior work that have high accuracy of detecting such attacks \cite{missing-data, shieldfs} but only a few of them utilize the ability of policy based allocation to recover from such attacks. The ones that utilize policy based allocation schemes \cite{afs, shieldfs} have significant overhead of read/write latencies, computation time, space requirement and limited recover-ability that are not feasible for Edge nodes. We discuss more limitations of such strategies in \textit{Section \ref{S:rw}} and identify the scope of improvement in terms of proper allocation with mechanisms allowing faster, portable and more efficient recovery. Our work primarily focuses on recovery in Hard Disk Drives (HDD), Flash Media and Solid State Devices (SSD).
%  Another major challenge that lies in such developments concerns the portability to previous platforms: how easily new systems can be integrated with pre-existing solutions \cite{portability}. Preference has always been given to those deployments that can be incorporated more easily than others, even if they do not lead to the most efficient solutions. 

Most of the conventional file systems such as Ext, NTFS and exFAT do not allow the users to specifically tag blocks/clusters/sectors - deleted or unused independently \cite{limitations-fs} where a block in hard-disk is a group of sectors that the operating system can point to. Lack of such freedom limits the kernel to overwrite the data on random locations reducing recovery. Some optimizations exist but are proprietary and not customizable to user specific needs. This limits the currently available kernels to utilize the full potential of file allocation and tagging for efficient recovery. Even the file allocation and mapping with virtual tables is restricted to specific fixed algorithms. Allowing these algorithms to be recover-ability aware in allocating blocks/clusters to files can improve the amount of data that can be recovered from such file systems. Making these algorithms adaptive and equipped with some learning model can further lead to optimization absent in current systems. 

% Another set of developments in the recovery domain have largely been for the recovery softwares without tinkering with the kernel and/or deeper levels \cite{recuva}. The file allocation policy allows file systems to allocate blocks such that later, when files are deleted, they can be recovered easily. This policy is also adaptive to the user's file access characteristics to further improve the recover-ability of file. It also opens new avenues for disk optimization at the level of recovery softwares. 
The proposed file system APEX (Adaptive Ext4), implements an adaptive file allocation policy that supports a wide diversity of platforms due to its portable implementation. It provides a significant improvement in recovery of files with low overheads. It is designed to be lightweight and easy to deploy in Edge/Fog computing frameworks, increasing their reliability and data protection. Another advantage is that it provides improved forensic based recovery for criminal investigations to expose evidence and hence catch hackers/invaders. The main \textbf{contributions} of our work are as follows:
\begin{itemize}
    \item We propose a lightweight, adaptive, portable and efficient file allocation system optimized for higher post deletion recovery which is flexible, robust and is independent of storage architecture
    \item We provide a set of pre-optimized weights that need only slight variation of hyper-parameters dependent on usage and thus low adapting time for new scenarios
    \item We develop a prototype file system APEX and show its efficacy on a real life scenario for malicious deletion/overwriting of video surveillance footage.
\end{itemize}
The rest of the paper is organized as follows. In \textit{Section \ref{S:rw}}, we provide related works and compare them with ours. In \textit{Section \ref{S:heuristics}} we first provide a basic recover-ability aware allocation mechanism and describe a heuristic based block ranking method that can optimize post-deletion data recovery. We then improve our heuristic measure by updating it dynamically to prioritize files based on the user's file access characteristics in \textit{Section \ref{S:model}} and also provide model level details of a Disk Simulator for learning the weights (hyper-parameters) of the block parameters, based on general-user file access characteristics. In \textit{Section \ref{S:implementation}}, we extend our implementation discussion to APEX file system using the FUSE (FileSystem in UserSpace) framework \cite{fuse}. In \textit{Section \ref{S:exp}} we provide a case study of overwriting video surveillance data using CryPy malware \cite{crypy} and provide experimental results of the model and comparison with other works both for recovery and overheads to show that APEX outperforms them. \textit{Section \ref{S:Conclusion}} concludes the paper and provides future directions to improve APEX.
%%%%%%%%%%%%%%%%%%%%%%%%%%%%%%%%%%%%%%%%%%%%%%%%%%%%%%%%%%%%%%%%%%%%%%%%%%%%%%%%
\section{Related work}
\label{S:rw}
\begin{table*}[ht]
\centering 
\resizebox{\textwidth}{!}{
\begin{tabular}{|c|c|c|c|c|c|c|c|c|c|c|c|}
\hline 
\multirow{2}{*}{Work} & Recovery specific & \multicolumn{4}{c|}{Low Overhead} & Selective files can & Allows custom & For & \multirow{2}{*}{Adaptive} & User & Cross \tabularnewline
\cline{3-6} 
 & allocation & Computation & Memory & Disk & File I/O & be marked critical & policies & Edge/Fog/Cloud &  &  specific & Platform \tabularnewline
\hline 
Ulrich et al. \cite{c3} &  &  &  &  &  &  &  &  & \checkmark &  & \tabularnewline
\hline 
AFS \cite{afs} & \checkmark &  &  &  &  & \checkmark & \checkmark &  &  &  & \tabularnewline
\hline 
Hanson \cite{c4} & \checkmark &  &  &  &  & \checkmark &  &  & \checkmark & \checkmark & \tabularnewline
\hline 
Breeuwsma et al. \cite{c5} &  & \checkmark &  &  &  &  &  &  &  &  & \checkmark \tabularnewline
\hline 
Alhussein et al \cite{ffs} & \checkmark & \checkmark & \checkmark & \checkmark &  & \checkmark &  & \checkmark &  &  & \tabularnewline
\hline 
Stoica et al. \cite{missing-data} &  & \checkmark & \checkmark &  &  &  &  &  & \checkmark &  & \tabularnewline
\hline 
Huang et al. \cite{flashguard} & \checkmark & \checkmark & \checkmark & \checkmark & \checkmark &  & \checkmark & \checkmark &  &  & \tabularnewline
\hline 
Alhussein et al \cite{c7} & \checkmark &  &  & \checkmark & \checkmark & \checkmark &  &  &  &  & \tabularnewline
\hline 
Baek et al. \cite{ssd-insider} &  & \checkmark & \checkmark &  &  &  &  & \checkmark &  &  & \checkmark \tabularnewline
\hline 
Lee et al. \cite{extsfr} & \checkmark & \checkmark & \checkmark & \checkmark &  & \checkmark &  & \checkmark &  &  & \tabularnewline
\hline 
Continella et al. \cite{shieldfs} & \checkmark & \checkmark & \checkmark & \checkmark & \checkmark &  &  &  & \checkmark &  & \tabularnewline
\hline 
APEX {[}this work{]} & \checkmark & \checkmark & \checkmark & \checkmark & \checkmark & \checkmark & \checkmark & \checkmark & \checkmark & \checkmark & \checkmark\tabularnewline
\hline 
\end{tabular}}
\captionof{table}{A comparison of related works with ours}
\label{tab:related-work}  
\end{table*}

The goal of data recovery is to recover lost data from a disk to maximum extent. This data might be `lost' because the disk has been corrupted or the files have been deleted. We focus on the latter aspect of the problem that too from data stealing/deleting malware. There has been significant work on data recovery at different levels, including file system, file allocation software, and recovery tools. However, there still lacks a holistic system that focuses on this aspect on generic, highly recoverable data storage by modifying the allocation policy. Table \ref{tab:related-work} provides a brief summary for comparing APEX with other systems.

There are two main directions in which the work on data recovery has progressed. The first concerns how data is recovered after deletion, and the second concerns with allocating data such that recovery later is optimized. Significant parameters for comparison include overheads, on demand recover-ability, adaptivity to different application scenarios, and custom policy employ-ability among others.
There has been work on dynamic file systems and allocation like by Ulrich et al. \cite{c3}, where the data is allocated between different drives for optimum resource utilization and distribution. Complex parity generation and comparison methods are used, spanning multiple drives, for improved utilization and recovery performance. However, it lacks priority based allocation and replacement policies that are optimized for recovery and access time. The popular Andrew File System (AFS) \cite{afs} in distributed systems also provides a backup mechanism to recover deleted or lost files for a limited period of time. This is not suitable for Fog nodes due to limited disk space available (there is high storage-to-compute cost ratio in Fog framework deployment and other communication limitations across Fog network). Many efforts have also been made in the directions of hardware tweaking and optimizations. An example is by Hanson \cite{c4}.

% Still, such optimizations fail to be incorporated inside current systems and lack model flexibility as per user needs. A lot of work has been done on forensic data recovery, for instance, by Breeuwsma et al. \cite{c5}. However, data has been recovered after the overwriting process, which limits the amount of recovery. Also, such methods are system specific (flash memories in this case).

Techniques that involve tagging of file blocks with some identifiers have been used by Alhussein et al. in \cite{ffs} and \cite{c7}, where frameworks like FUSE have been used for development of forensic based file-systems. They provide forensic file identifiers at cluster-level file allocation to provide information needed for file types to be recovered after deletion.  As they are only limited to file cluster identification and the identification of file types, the amount of recover-ability of data is limited because they completely ignore the file usage characteristics and temporal locality. Another adaptive approach by Stoica et al. \cite{missing-data} uses weighted least squares Iterative Adaptive Approach (IAA) to detect and recover missing data, but this works mostly across streams of data signals and is not scalable to actual file systems. We consider it because of its unique adaptive approach to categorize cluster cells for efficiently allocating them in buffers/disk. Other works like by Continella et al. \cite{shieldfs} or Baek et al. \cite{ssd-insider} provide a self-healing ransomware-aware filesystem which is capable of both detection and  the recovery from ransomware attacks. It works by analyzing the I/O data trace of various processes and uses a classifier to detect if the process is maliciously deleting data. If such a process is discovered, it uses a recovery approach used by copy-on-write filesystems. This poses significant overheads in terms of disk space and I/O bandwidth requirements and hence is not optimum for Edge nodes. Another problem is that, it has non-zero detection error in file destructive ransomwares as it has been trained for those that are encryption based. We tested it for Gandcrab and it was not able to identify it. 

% Similar is the case with the work of Baek et al. \cite{ssd-insider} as they propose an internal defense mechanism for Solid State Drive (SSD) against ransomware. Also, like \cite{afs}, they use delayed deletion which increases storage requirements and hence not suitable for Fog computing domains. 
Another technique by Huang et al. \cite{flashguard} has been proposed for recovering encrypted data from flash drives. They provide a ransomware tolerant SSD which has firmware-level recovery systems, but the main drawback of their approach is that they keep multiple copies of data and hence need significantly more space than required. Another disadvantage is that their approach is specific to SSDs and not for generic storage devices. Lee et al. \cite{extsfr} propose ExtSFR, a scalable file recovery framework for distributed filesystems in IoT domains. This uses files' metadata to identify and recover them, but ignore file usage and access characteristics which limit recoverability. In aforementioned works, file access characteristics, recovery based allocation strategies, and cluster identifiers are exploited from a narrow perspective and capabilities of adaptive priority based allocation approaches have not been fully leveraged. In addition, most systems overlook the constraints in edge paradigm and hence are not cost and/or energy efficient for such deployments. APEX ensures data security using adaptive prioritization of blocks in storage media based on recovery heuristics and provides efficient mechanisms that have minimal compute, bandwidth and space overheads.

\section{Recoverability Aware File allocation}
\label{S:heuristics}

In the previous section, we described the flaws of existing systems and emphasized on how a recovery-aware file allocation system can provide a more robust and efficient mechanism for recovering deleted data. Here we provide the details of implementing such system and later describe how much improvement in terms of recovery it provides. For this work, we have kept the threat model as a malicious entity which directly attacks the system to sabotage critical files while the attack surface is limited to the user application wherein the kernel is secured.

% \subsection{A Brute-Force based model}

% A simple allocation mechanism for recoverability-aware file allocation could have a performance metric like the amount of data recovered. Based on this metric, whenever we need to allocate a file of \textit{k} blocks to \textit{n} unused blocks in disk, we can try all possible $\Comb{n}{k}$ combinations, compare the size of data recovered and choose the set of \textit{k} blocks with maximum such size. Though this is a correct approach, it is not scalable due to exponential time complexity. 

% \subsection{Heuristics based simplification}

% To have a scalable approach that is recoverability-aware and has low file access overhead, we use a ranking procedure to prioritize blocks. This priority based ranking enforces that the blocks which are more critical for the user are either not overwritten or are overwritten at the end. The APEX file system uses a model based on such prioritization. To prioritize blocks in a way that the highest priority one has least effect on recovery, we propose and reason the importance of four major block/file level parameters that act as heuristics. These parameters are combined into a prioritization score called the Priority Factor ($PF$).

\subsection{Block Parameters and Priority Factor}
The four proposed parameters that act as heuristics to rank a block for its priority for allocation are History Factor ($HF$), Usage Factor ($UF$), Spatial Factor ($SF$) and Linking Factor ($LF$). The Priority Factor ($PF$) of a block is a linear combination of these parameters. The weights of theses factors are kept as hyper-parameters which are dynamically learned for improved recovery based on the user's file usage characteristics in the APEX model. This priority score is used to sort the unused blocks in a priority queue which then is used for finding fresh blocks to allocate to a new file. Now, the Priority factor ($PF$) is defined according to the equation: 
\begin{equation*}
PF \ = \ \lambda \cdot HF \ - \ \sigma \cdot UF \ + \ \rho \cdot SF \ + \ \mu \cdot LF 
\end{equation*}

Priority Factor is periodically calculated for each unused block of the disk. Here the hyper-parameters: $\lambda,\ \sigma,\ \rho,\ \mu$ which are coefficients of the disk block parameters. They are dynamically updated based on the user's file access characteristics. The different types of clusters/blocks are \textit{Used} and \textit{Unused}. When a file is created, some unused blocks are allocated and thus belong to the \textit{Used} category. When a file is deleted, the \textit{Used} blocks that correspond to that file are converted to \textit{Unused} category. The different types of files in APEX nomenclature are:
\begin{itemize}
    \item Used: File exists in disk
    \item Deleted: File exists in disk but deleted and blocks can be overwritten (only partial fragments of the file may be present)
    \item Obsolete: No block of the file exists in the current state of the disk
\end{itemize}
Different types of operations are read, write, delete, create (new file). We now define the block parameters and reason their importance:

\subsubsection{History factor ($HF$)} The history factor accounts for how old a particular file's blocks are, in terms of "delete" or "over-write" operations. Each time a file's blocks are deleted or over-written, the rest of the blocks' $HF$ increases by one. At block level the history factor can be visualized as shown in Figure \ref{Fig:hf}.
% \begin{figure}[h]
% \centering
% \includegraphics[width=5.5cm]{graphs/HF}
% \caption{History factor of a block}
% \label{Fig:hf}
% \end{figure}
Here, the "file" of the block shown as "a" represents the set of blocks that belong to the last file of which block "a" was part of. For different cases of a block, we have:
\begin{itemize}
\item Unused: for every Delete/over-write operation, the HF is raised by 1
\item Unused to used transition: set HF to 1
\item Used: no change
\item Used to unused transition: set HF to 0
\end{itemize}
This parameter exists for each block in the disk. The reason HF is important for recovery is: if blocks of the same file are overwritten then the extent of recovery of other blocks decreases. This is also mentioned in other works \cite{afs, ssd-insider, hf-1} where they use similar notions to capture the history of file. When HF increases, it makes the recovery of the cluster more difficult and hence PF should increase, which implies that the positive scalar constant ($\lambda$) will be positive.

\subsubsection{Usage Factor ($UF$)} 
\begin{figure}[t]
\centering
\subfigure[History Factor]{\includegraphics[width=0.45\linewidth]{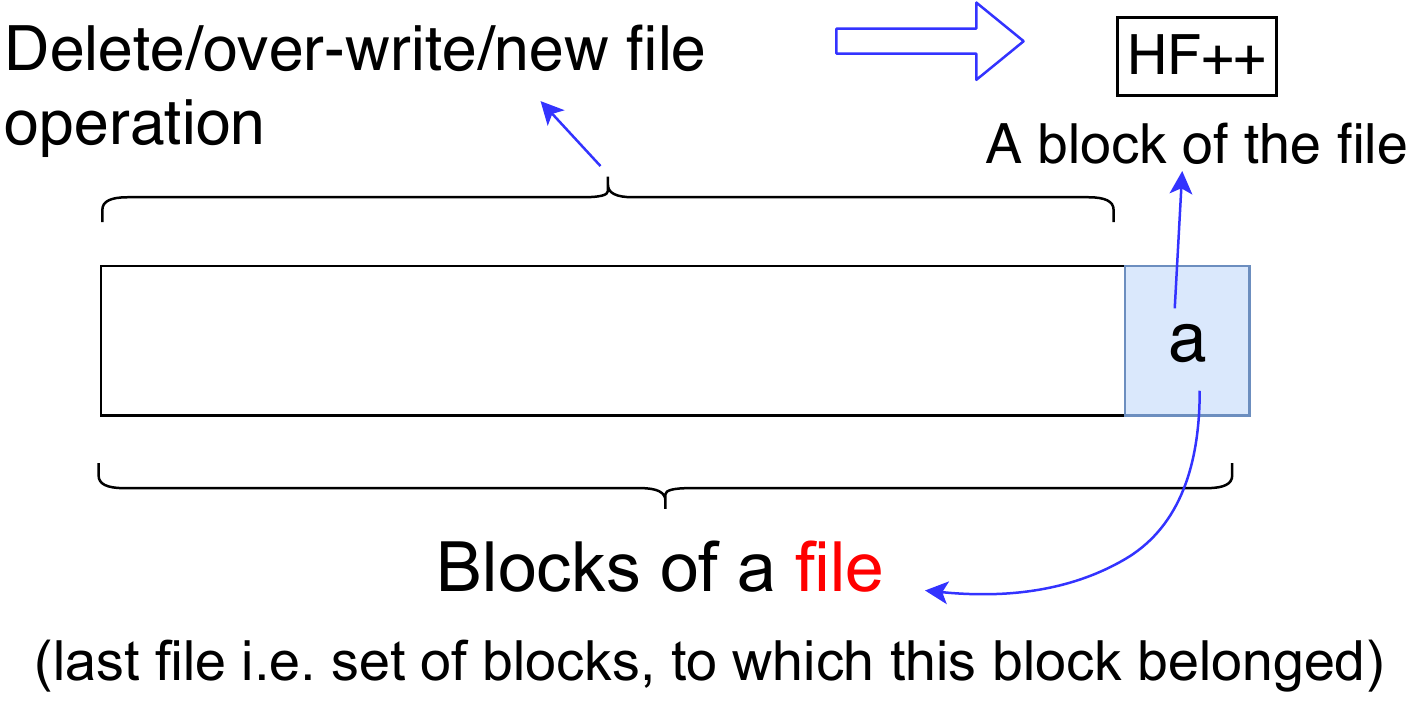} \label{Fig:hf}}
\subfigure[Usage Factor]{\includegraphics[width=0.44\linewidth]{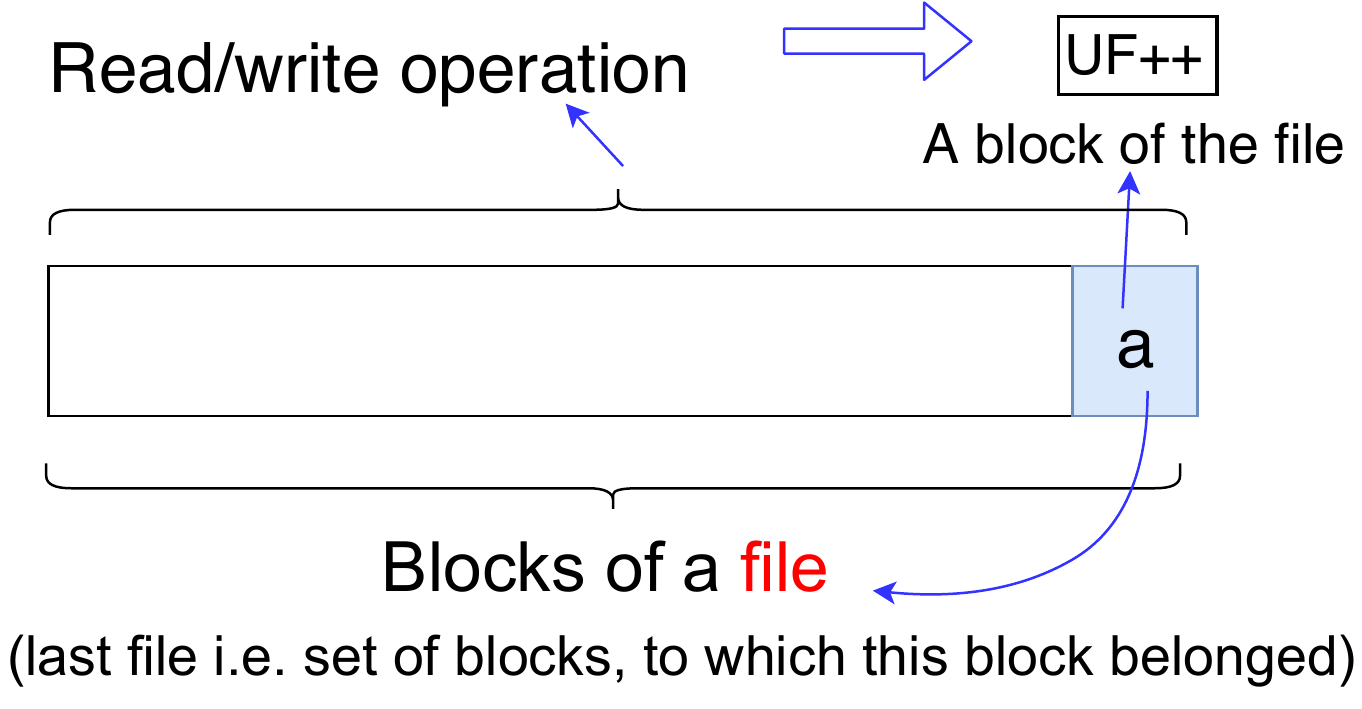} \label{Fig:uf}}
\caption{Block heurisics}
\label{Fig:heuristics}
\end{figure}

The usage factor takes into account the usage of a file (and thus its blocks). It is quantified by the number of read/write operations on that particular file before deletion/over-writing. The higher the usage of a particular file, the more recovery sensitive (important to user) the file gets. The change to the usage factor based on the file operation is shown in Figure \ref{Fig:uf}. The UF of all blocks in a disk are initialized to 0.\\
For different cases of a blocks, we have:
\begin{itemize}
\item Used: For each read/write operation of any blocks of the file, the UF of each file increases by 1
\item Used to Unused transition: no change
\item Unused: no change
\item Unused to used transition: set UF to 1 when a new file is created
\end{itemize}
Other works \cite{ssd-insider, hf-1} also use similar notions of file/block prioritization based on usage of file. This is because frequently used files are considered more important to the user. This means that the PF (priority for overwriting it) should reduce and hence the scalar constant ($\sigma$) has negative sign with it.

% \begin{figure}[h]
% \centering
% \includegraphics[width=4cm]{graphs/SF}
% \caption{Spatial factor of a cluster}
% \label{fig:sf}
% \end{figure}

\subsubsection{Spatial Factor ($SF$)}   The spatial factor includes the possibility of the recovery of blocks located especially in the neighborhood of a particular block. Its importance has been shown in other works as well \cite{ssd-insider, sf-1} because of its direct effect on file access time. The spatial factor of a block is higher if the overall priority factor of the neighboring blocks are high. Thus, SF is kept as the average of blocks that are physically spatially adjacent. `Spatial Adjacency' depends on the physical characteristics of the medium. For HDD we may consider the blocks in the same sector as the neighboring ones.
As the blocks being allocated for a new file are the high priority blocks, if the neighboring blocks have high PF, the spatial locality would increase. Consequently, this block should also get replaced and hence for high SF, the PF should be high. This shows that the scalar constant ($\rho$) is positive. SFs of all blocks are updated after each I/O operation.
For different cases for a block, we have:
\begin{itemize}
\item Used: Reset to 0
\item Unused: Average $PF$ of nearby blocks
\end{itemize}
This factor is initialized to 0. This would be very useful in HDDs, and our algorithm can take into account on the fly de-fragmentation in such storage devices. However, the introduction of $SF$ should not hamper with the wear leveling algorithms employed by flash drives and thus this factor is dropped in random access drives. For random access drives and SSDs, optimizing spatial locality is not required.

\subsubsection{Linking Factor ($LF$)} The linking factor depends on the format of a particular file and captures the extent of recovery possible for specific file formats. Some files like .jpg, .mp3, .avi, can be partially recovered even if some blocks are deleted/over-written. Such type of files should be distinguished from other ELF (Executable and Linkable Format) types like .exe which can not be recovered even when only one block has been over-written \cite{lf-1}. Blocks that are unused but the last file to which they belonged are present and the file belongs to the non ELF class of formats, would have $LF$ = 0. Others like those which belonged to file with ELF formats would have $LF$ = 1. This factor is initialized to 1, as in the beginning these files are not linked to anything. The scalar constant ($\mu$) should be positive in this case as well. The Priority Factor ($PF$) depicts the priority of a block to be overwritten by new data. The higher $PF$ blocks would be ready for over-writing first. The blocks with low $PF$ are more sensitive to recovery.

\subsection{Functional Model Working}

We use $PF$ as a reasonable heuristic to rank blocks in the decreasing priority of being allocated to new files. Block allocation, whenever a new file is created, is thus based on the priority factor. The blocks with the highest priority factors are allocated to the new file, number equal to those required for the new file. Whenever a file is deleted, its used blocks are shifted to the unused blocks' set and this set is again used for allocation to new files with allocating those blocks first that have the highest PF. 

We now show how the ranking based on priority factor can be improved further by tuning the hyperparameters: $\lambda,\ \sigma,\ \rho,\ \mu$ dynamically. This tuning is based on the file access characteristics of the user to allow more efficient allocation.

% \begin{figure}[h]
% \centering
% \includegraphics[width=9cm]{graphs/mdp}
% \caption{One dimensional MDP showing state transition for one variable}
% \label{Fig:MDP}
% \end{figure}

\begin{figure}[b]
\centering
\includegraphics[width=0.9\linewidth]{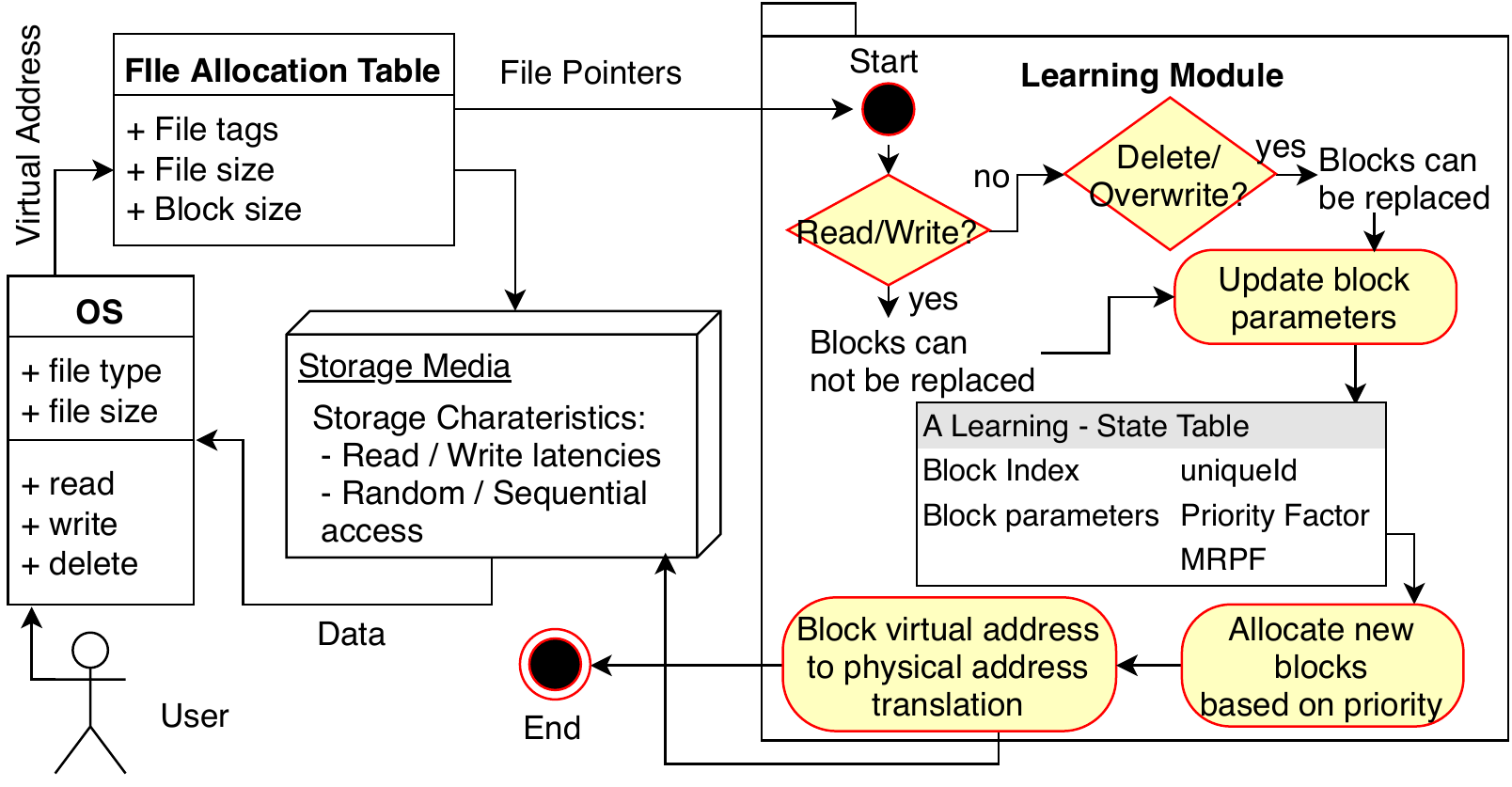}
\caption{Model description using UML (Unified Modeling Language)}
\label{Fig:UML}
\end{figure}

\section{Hyper-parameter optimization: Adaptive allocation model}
\label{S:model}
In the previous section, we described how priority factor can be used to rank blocks to improve post-deletion data recovery. This heuristic measure is still an approximation and there exists scope for further tuning the hyperparameters: $\lambda, \sigma, \rho, \mu$ discussed before. This tuning can be dynamic based on user's file access characteristics which can make this heuristic more precise and lead to better allocation. We next describe a simulated environment close to real life scenario to determine the optimum values of the hyperparameters.

\subsection{Learning Model}

We use $Q$-Learning, which is a reinforcement learning model to optimize the set of hyper-parameters. Here, the state of the $Q$-Learning Model is defined by the tuple of hyper-parameters. An action is defined by the increment/decrement of the hyper parameters. Converging to the optimal set of these hyper parameters would be the goal of the model. The model is divided into two parts, the first is the learning model which optimizes these parameters. The other part is the allocation system, which allocates blocks to new files based on the current hyper-parameter set and corresponding $PF$s of the learning model.

Considering each I/O operation as one iteration, the learning model updates the coefficients at each iteration and gradually converges to the best set of values. The coefficients are updated frequently because of the dynamism of disk/file accesses. The model would be trained on the basis of a Performance measure ($P$), which is a combination of the average recovery ratio of files and the spatial locality (only recovery ratio for random access drives). Each action would change this measure, and by the learning algorithm the model would converge to a optimum hyper-parameter set. In the building of the prototype File System APEX, we have trained the $Q$ learning model based on some representative file access rates and distribution of common file types.

Figure \ref{Fig:UML} shows a diagrammatic representation of the model and how the learning model interacts with the File Allocation Table (FAT), OS and the disk. As the state space needs to be finite and the model should be able to converge to a definite solution, the set of values on which the hyper-parameters vary is kept as: {1,2,..,10}. This range is based on empirical results and convergence constraints. This gives a crude idea as to what are the values required for the best recovery efficiency keeping the least count of value measures to be 1. 

\subsection{Objective Function}
The objective function, referred as performance $P$ of the learning model is implemented as a linear combination of the weighted average Recovery Ratio (RR). RR is a standard metric for comparing recovery of file systems \cite{shieldfs, ffs} and is weighted by the usage frequency of the files. Moreover, a measure to quantify average access time of files in the disk \cite{ssd-insider, sf-1} is also common metric. We define for each file:
\begin{itemize}
\item \textit{Recovery Ratio} (RR): 
\begin{itemize}
\item For executable, objects and archived file types: 1 for complete recovery, 0 for partial/no recovery.
\item For files including text, word processor, multimedia, .pdf etc.: $ \frac{Data\ recovered\ in\ bytes}{Original\ file\ size\ in\ bytes} $ if meta-data recovered else, 0. (Because file cannot be even partially read without the meta-information)
\end{itemize}
\item \textit{Approximate Access Time} (AAT): Approximate measure of access time of a file which is the time for last read/write operation. When a file is created it is the file creation time.
\end{itemize}
Now, Performance P is defined as a convex combination of the above terms: \\
\begin{center}
$ P = \alpha \frac{100 \, \cdot \, \sum_{all\, deleted\, files} RR\cdot UF}{\sum_{all\, deleted\, files}UF} \, - \, \beta \frac{\sum_{all\, current\, files} AAT}{Number\, of\, current \, files}$
\end{center}
Here, $0 \leq \alpha \leq 1$, $0 \leq \beta \leq 1$ and $\alpha + \beta = 1$. Both terms of which $\alpha$ and $\beta$ are coefficients, are dependent on each other. The term involving access time depends not only on the spatial distribution of files but also the average size of current files which affects recovery of other files. It also affects recovery because in many spatial distributions the meta-data is erased which makes a partially recoverable file (i.e. some blocks of data exist) have RR = 0. The other term that involves recovery ratio depends on how many blocks of deleted files are present and also on which of them are present (meta-data or data). Different values of $\alpha$ and $\beta$ are applicable to different scenarios. If we want to optimize recovery only, then $\alpha = 1$ and $\beta = 0$, for example, where edge devices store critical data like video surveillance footage or health monitoring data. Such applications do not require high I/O bandwidth but critically require recoverability. If we want to optimize I/O only then $\alpha = 0$ and $\beta = 1$. This case would arise in edge configurations with high bandwidth data streams and all data being stored in a separate database node or cloud which requires fast I/O.

\subsection{Disk Simulator}

% The Disk simulator works at the physical block allocation level for files. This simulator creates a disk (non-physical) that works at block levels to generate factor signatures as "parameter maps". Such maps are the parameter values for each block in disk and could be created for each parameter.
% \begin{figure}[h]
% \centering
% \includegraphics[width=4cm]{graphs/parameter_map}
% \caption{Parameter map of a state factor}
% \end{figure}
The disk block size for simulation model is kept the same as the real value = 4KB. Total disk space is kept as 256MB and a file can vary from 16KB (4 blocks) to 1024KB/1MB (256 blocks). Based on the rules for parameter update, the parameter maps are created for each state, i.e, a hyper-parameter set. This simulator model also allocates the blocks (based on ranking on $PF$) whenever there is a call from OS simulator (explained next). It also keeps track of the MRPF (Most Recent Parent File) for each block, which is a pair of the filename that a specific block belongs to (or did before logical deletion) and the set of blocks that were allocated to this particular file. 

\subsection{OS / File IO Simulator}

This simulator does the file level/ OS level disk management. This introduces random file operations which belong to one of these categories: (1) Read/write, (2) Create file, (3) Delete file. The "Create file" operation updates the MRPF of each block that belongs to that new file after the operation. 

\subsection{Learning Hyper Parameters}

\begin{figure}[t]
\centering
\includegraphics[width=0.8\linewidth]{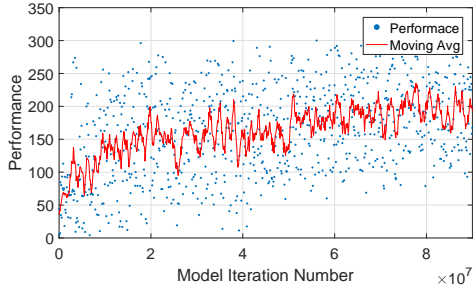}
\caption{Performance with Model Iteration Number (MIN)}
\label{Fig:perf}
\end{figure}

Based on the definition of the hyper-parameters and performance function, the optimization problem is formulated as:
\begin{equation*}
\begin{aligned}
& \underset{\lambda, \sigma, \rho, \mu}{\text{maximize}}
& & P \\
& \text{subject to}
& & \lambda, \sigma, \rho, \mu \in [1,10] ;\ \lambda, \sigma, \rho, \mu \in \mathbb{N}, \\
&&& Blocks\ allocated\ to\ new\ files\ in\ decreasing\ \\
&&& order\ of\ PF  = \lambda HF - \sigma  UF  +  \rho  SF  +  \mu  LF 
\end{aligned}
\end{equation*}
The model has been simulated using two iteration counts:
\begin{enumerate}
\item Model Iteration Number (MIN): This increments by one for each state change (action) in the learning model. 
\item Operation Iteration Number (OIN): This increments by one for each File I/O Operation performed by the OS Simulator
\end{enumerate}
The File I/O operations should be greater than the model actions as it is not much useful to have a learning model that analyses the state space after many operations. To learn the optimal values of the hyper parameters, we set the MIN increment after 1000 OIN increments. This helps to reach a stable configuration after many I/O operations and in evaluating the disk performance based on the current set of hyper-parameter values.
% The performance graph generated from the simulation showing $P$ with MIN is shown in Figure \ref{Fig:perf}. 
The conditions kept for the learning stage are:
\begin{enumerate}
    \item Disk Size: 16 $\times$ 16 = 256 blocks
    \item Maximum file size = 20 blocks
    \item Percentage of linked files = 20 (stochastically varying around this value)
    \item Minimum disk Utilization = 70\%
\end{enumerate}
As in general $Q$-Learning models, this learning model has an "Exploration factor" depicted as $\epsilon$. This factor drops down from 1 exponentially. It decides with what probability a random action takes place. The probability with which a random action is chosen is $\epsilon$ and for an optimal action it is $1 \, - \, \epsilon$. As $\epsilon$ decreases, the model chooses the action with the highest $\Delta P$, which is the change in the performance function of the learning model. Gradually, as the model explores, $\epsilon$ reduces and model converges to the optimal state. For this experiment, the learning model converges when $\epsilon$ is $\leq$ $3\times10^{-5}$. For the simulation setting at this convergence point: MIN reaches $9.01\times 10^7$ and OIN reaches $9.01 \times 10^{10}$. The performance measure, as described earlier, starts from 44.00 (based on Ext4 prioritization factors). The value of P tends nearly 190 with time, thus based on the equation of performance, there is an approximately 280\% improvement as shown in Figure \ref{Fig:perf}. The hyper-parameter values converge to (4, 7, 1, 9).

% \textbf{Sanity Check}: The converged values show that coefficients of the different factors (HF, UF, SF, LF) considered in the model, would be (4, 7, 1, 9) for maximum recovery ratio and minimum average access time. The SF has the least contribution which is reasonable because in the model, we have assumed the AAT to be linear function of number of file blocks and it does not depend on spatial distribution of blocks (in random access devices). The coefficient of HF is moderate as expected, because of the contribution of the file modification history as compared to Linking Factor importance. It is higher than that of SF as it accounts to the recovery ratio based on how many delete/overwrite operations have been performed on remaining blocks of the file. LF has the highest contribution which is also reasonable, because if LF is 0 then recovery is of no use at all. Since the LF value (0 or 1) is a Bernoulli trial, there are on an average 20\% the number of blocks, with very less recovery scope (like .exe files, LF = 1). This has a huge impact on recoverability, and thus a high coefficient. UF has high contribution which from the initial experiments shown in \textit{Section \ref{S:motivation}} leads to a large impact on performance. As the number of Read/Write operations is less than general, $\sigma$ is high. Files with high UF are more critical for recovery.

This set of hyper-parameter values are then used in a dynamic setting for block prioritization in the APEX framework. The initial Priority Factor, which is updated on-the-fly based on the learning model, used for the block allocation for new files in APEX is:
\begin{align*}
PF \ &= \ 4 \cdot HF \ - \ 7 \cdot UF \ + \  SF \ + \ 9 \cdot LF
\end{align*}

% \begin{figure*}[ht]
% \centering
% \includegraphics[width=16.5cm]{graphs/uml.png}
% \caption{Class Diagram of the Disk Simulator framework}\label{Fig:uml-APEX}
% \end{figure*}

%%%%%%%%%%%%%%%%%%%%%%%%%%%%%%%%%%%%%%%%%%%%%%%%%%%%%%%%%%%%%%%%%%%%%%%%%%%%%%

\section{Implementation of APEX using FUSE}
\label{S:implementation}
To implement APEX, we used the FUSE framework \cite{fuse}. FUSE (Filesystem in Userspace) is an interface for user programs to export a file-system to the Linux kernel. FUSE mainly consists of two components: the FUSE kernel module and the lib-fuse library. Libfuse provides the reference implementation for communicating with the FUSE kernel module. It is the most widely used library with active support that allows users to develop, simulate and test their own file-system without writing kernel-level code. 

% There exist wrappers for multiple programming languages including C, C++, and JAVA. which provide easier implementations and libraries for supporting data structures that we intend to use.

To implement APEX, we need the following structures: (1) Disk, (2) Directory, (3) File, (4) Block. For the Disk simulation system, we only required the \textit{Disk}, \textit{File} and \textit{Block} structures while modifying them according to FUSE format and include a new \textit{Directory} implementation for APEX. Instead of \textit{Disk} being a 2D array of Blocks, now its a large \textit{ByteBuffer} which is indexed by \textit{Blocks}. Disk contains the optimal coefficient values of the parameters (found from the RL model), current and deleted \textit{file} list a \textbf{Heap} of unused blocks and a \textbf{HashSet} of used blocks. The \textit{File} and \textit{Directory} are derived from an abstract class \textit{Path} which has specifications required for a File system structure. \textit{Directory} can have several children \textit{Paths} (which can be a file or directory) recursively. \textit{File} has several children \textit{Blocks}, file-type (FUSE \& linking factor).

At the block level, main operations are:
\begin{enumerate}
    \item \textbf{Allocate}: Sets the block's state to used, initializes the block's parameters, updates the parent file pointer and obtains the linking factor (found by the parent file's \textsc{FuseFileInfo}).
    \item \textbf{De-allocate}: Sets the state to unused, updates the block's parameters and parent pointer.
    \item \textbf{Write}: Writes the required content to the block. Includes writing from a buffer and offset.
    \item \textbf{Read}: Returns the block's data.
    \item \textbf{Change}: Updates the individual factors of the block.
    \item \textbf{Update}: Updates the priority score of the block and changes its position in the unused heap (if applicable).
\end{enumerate}

\section{Overwriting video Surveillance using CryPy malware: A Case Study}
\label{S:exp}

In order to compare APEX with other works, we use a Fog environment built on FogBus \cite{FogBus} for video surveillance on Raspberry Pi based Fog Nodes. As such nodes are usually connected to cloud via the Internet, many viruses, trojans, etc. are likely to creep into the network. We compare APEX with other recovery specific systems that can be integrated in Edge nodes: AFS \cite{afs}, FFS \cite{ffs}, ShieldFS \cite{shieldfs}, ExtSFR \cite{extsfr}.

\subsection{Setup}

The machine - Raspberry Pi 3: Model B, used for the experiments has the following specifications:
\begin{enumerate}
    \item SoC Broadcom BCM2837
    \item CPU: 1.2 GHZ quad-core ARM Cortex A53
    \item RAM: 1 GB LPDDR2-900 SDRAM
    \item Storage: 100 GB WD Black HDD or Samsung T5 SSD
\end{enumerate}

\subsection{Experiment 1: Recovery ratio performance}
\label{S:rr}
In order to compare the data recovery performance, similar tests were executed on the APEX File System and the Base (Ext4) File System, with other works and recovery ratios were compared. For a simple analysis, we keep the file system size for user data limited to 1GB. First, primary data was deleted using the CryPy malware \cite{crypy} after custom modifications, and secondary data was written on the drive. After this, the primary data was recovered. We perform tests for both HDD and SSD, which give very similar results (difference of $<0.05$ recovery ratio) and hence report the average.
\subsubsection{Test files}
Primary Data was kept consistent throughout the experiment to maintain consistency and consist of a set of 5 sample videos each of size = 95 MB. Secondary Data size was increased in each iteration by predefined constants and only written after soft deletion of primary data. Data recovered was measured in terms of the size of recovered primary data. Also measured in terms of recovered files that were view-able after recovery

\subsubsection{Recovery ratio}$Data \, recovered\, /\, Primary\, Data\, size$

\subsubsection{Observations}

% \begin{center}
% \begin{tikzpicture}[scale=0.85][!b]
% \begin{axis}[
%     title={Recovery ratio with Secondary data size},
%     xlabel={Secondary Data Size [MB]},
%     ylabel={Recovery Ratio},
%     xmin=0, xmax=900,
%     ymin=0, ymax=1.2,
%     xtick={34, 129, 224, 319, 414, 509, 604, 699, 794, 889},
%     ytick={0.1,0.2,0.3,0.4,0.5,0.6,0.7,0.8,0.9,1.0,1.1},
%     legend pos=north west,
%     ymajorgrids=true,
%     legend style={at={(1,1)},
% 	anchor=north east,legend columns=-1},
%     xmajorgrids=true,
%     grid style=dashed,
% ]
% \addplot[
%     color=blue,
%     mark=square,
%     ]
%     coordinates {
% (34,1)(129,1)(224,1)(319,1)(414,1)(509,1)(604,0.884)(699,0.684)(794,0.484)(889,0.284)
%     };
% \addplot[
%     color=red,
%     mark=square,
%     ]
%     coordinates{
% (34,0.928)(129,0.728)(224,0.528)(319,0.328)(414,0.128)(509,0)(604,0)(699,0)(794,0)(889,0)
%     };
%     \legend{APEX, Ext4}

% \end{axis}
% \end{tikzpicture}
% \captionof{figure}{Recovery ratio for different secondary data sizes}
% \label{Fig:rr}
% \end{center}

\begin{figure}[t]
\centering
\subfigure[Secondary data size = 414 MB]{\includegraphics[width=0.48\linewidth]{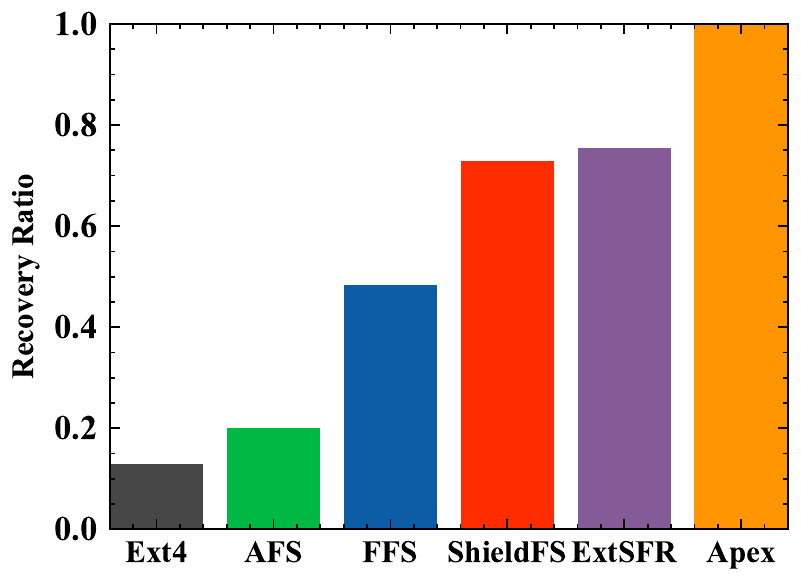} \label{fig:recovery1}}
\subfigure[Secondary data size = 794 MB]{\includegraphics[width=0.48\linewidth]{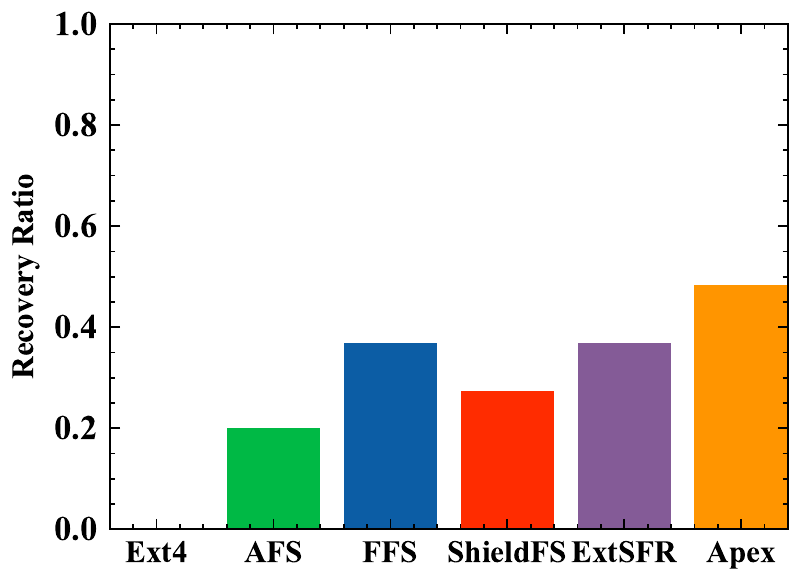} \label{fig:recovery2}}
\caption{Recovery ratio for different Secondary data size}
\label{Fig:recovery3}
\end{figure}

Across different file systems as shown in Figure \ref{fig:recovery1} and Figure \ref{fig:recovery2}, for the Secondary data size = 414 MB and 794 MB respectively, we see that the APEX file system has the highest recovery ratio. As there is no policy for recovery in Ext4, it has the least (even 0 MB in second case) recovery. As the surveillance footage was taken such that each video accounts for 1 hour of data, AFS file system with 1 hour of delayed deletion causes 4 out of 5 files to be permanently deleted. This remains unchanged in both experiments. For others including FFS, ShieldFS, and ExtSFR; as they use only file identifiers in blocks and not which files were most recently and frequently used (video files in our case), they overwrite their data and hence lead to lower recovery. APEX on the other hand allocates new files i.e. Secondary data to separate locations preventing overwriting of Original data, is able to recover maximum. APEX system improve the recovery ratio by 678\%, and 32\% compared to the base Ext4 and ExtSFR respectively for 414 MB secondary data and 31\% compared to ExtSFR for 794 MB secondary data.

% \begin{figure}[b]
% \centering
% \includegraphics[width=6.5cm]{graphs/write.pdf}
% \caption{Write speed in KB per second}
% \label{Fig:bench1}
% \centering
% \includegraphics[width=6.5cm]{graphs/read.pdf}
% \caption{Read speed in KB per second}
% \label{Fig:bench2}
% % \end{figure}
% % \begin{figure}[ht]
% \centering
% \includegraphics[width=6.5cm]{graphs/delete.pdf}
% \caption{Delete speed in KB per second}
% \label{Fig:bench3}
% \end{figure}

\subsection{Experiment 2: Read-write performance}
\label{S:read-write-experiments}

As the APEX file system adds the block level - prioritized file allocation implementation over Ext4, which leads to additional overhead, thus the FUSE implementation of APEX was compared to the Base file system, which in our case is \textit{Ext4}. To test the read, write and delete speed in both systems, we used FileBench Filesystem benchmarking tool \cite{c8}. These benchmarks provide read, write and delete speeds for \textit{n} files, each of size \textit{m} (where \textit{n} and \textit{m} are given as inputs to the benchmark code). For the current study, random files were sequentially created, read, written and deleted where \textit{n} varied from 10 to 1000 and \textit{m} varied from 1 MB to 1000 MB. %Both APEX and exFAT were formatted with a file system having 1MB of usable storage.\\

\begin{figure}[t]
\centering
\subfigure[Write speed in KB/s]{\includegraphics[height=0.37\linewidth]{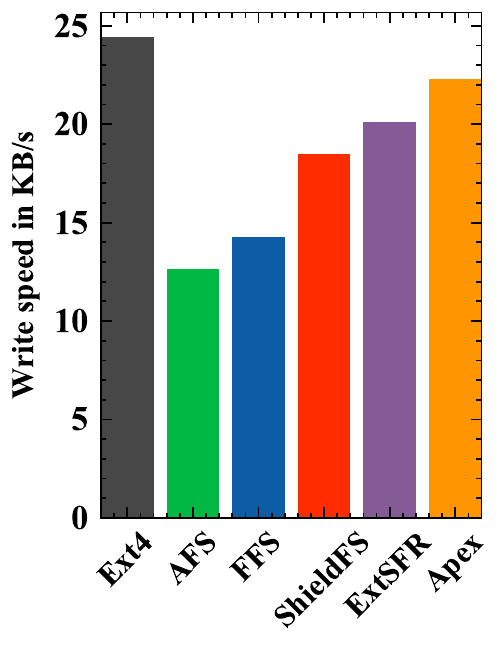} \label{Fig:bench1}}
\subfigure[Read speed in MB/s]{\includegraphics[height=0.37\linewidth]{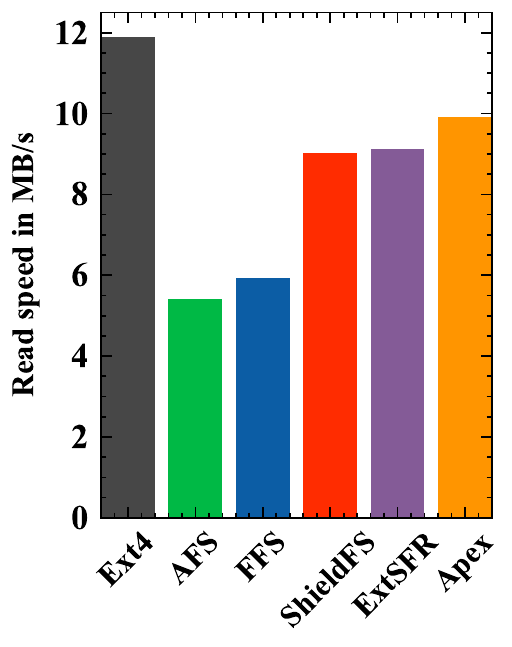} \label{Fig:bench2}}
\subfigure[Delete speed in MB/s]{\includegraphics[height=0.37\linewidth]{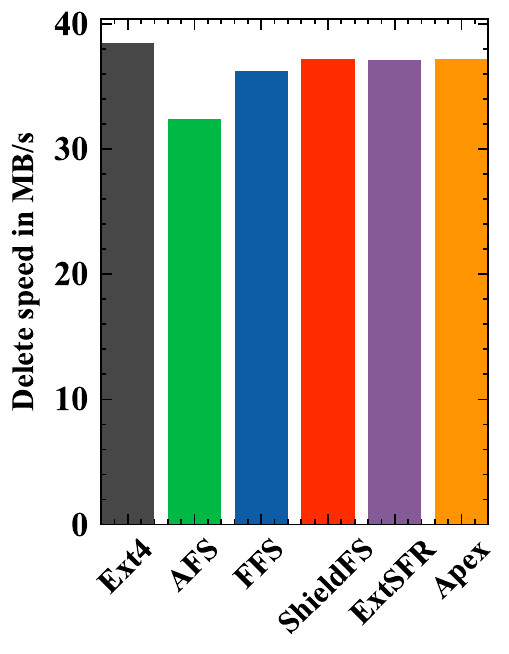} \label{Fig:bench3}}
\caption{Write, Read and Delete speed comparison}
\label{Fig:recovery}
\end{figure}

Figures \ref{Fig:bench1}-\ref{Fig:bench3} show the results in bytes per second for writing, reading and deleting for different frameworks using HDD media (which has higher overhead so we report these results). The graphs show that the delete operation speed for all frameworks except AFS is close to Ext4. Read/write speeds are lower than Ext4 but the overhead is minimum in APEX. This is primarily because of identifier mapping in blocks being updated periodically in FFS. In ShieldFS, the detection algorithm has its overhead which continuously monitors the disk footprint of filebench program and consumes bandwidth. Overall, APEX has the least overhead among all implementations.

\subsection{Experiment 3: CPU and RAM overhead}
\label{S:cpu-ram}

\begin{figure}[t]
\centering
\subfigure[CPU usage (\%) in different frameworks]{\includegraphics[width=0.45\linewidth]{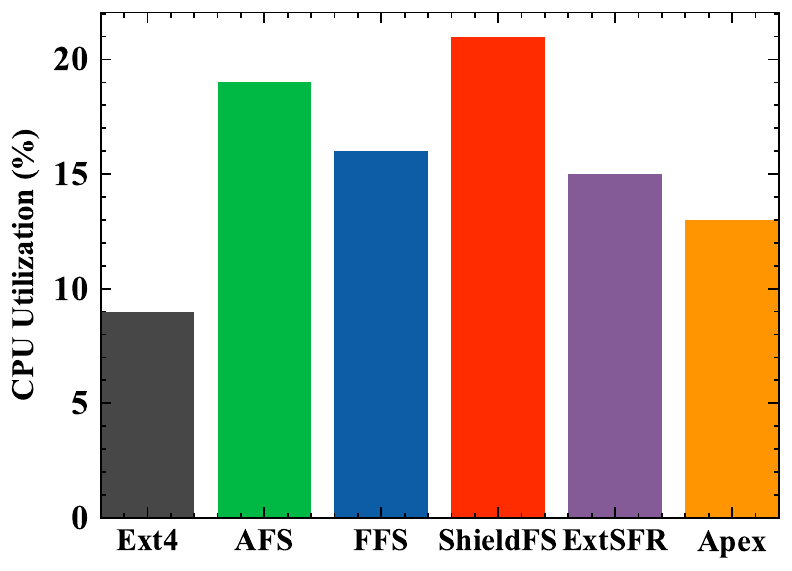} \label{Fig:cpu}}
\subfigure[RAM usage (\%) in different frameworks]{\includegraphics[width=0.45\linewidth]{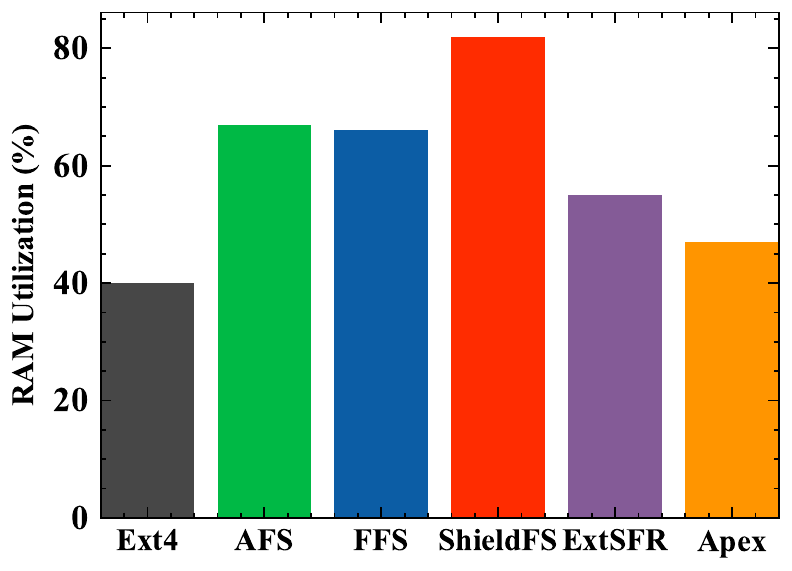} \label{Fig:ram}}
\caption{CPU and RAM comparison}
\label{Fig:cpu-ram}
\end{figure}

Due to various complex background tasks of process I/O detection in ShieldFS, file block identification in FFS and backup of data in AFS, the CPU and RAM usage of such frameworks is much higher compared to APEX. ShieldFS uses many sophisticated cryptographic primitives to identify if a program could be `potentially malicious'. This requires constant overhead of computation and memory. ExtSFR has a post-deletion inode tracing and journal checking protocol which needs to check the whole file system for very small changes. APEX maintains the state of the files, hence only the files of interest can be checked. FFS maintains `forensic file system identifier' for each file and identifies the relevant information needed for recovery using such identifiers. These identifiers are stored in disk and are required to be fetched when checking for recoverable files in the file system, which makes it slow. APEX uses the disk cache information to update the block parameters of the most used blocks more frequently than others. This allows APEX to have a very small working set at any instant of time (maximum 7\% in current tests). It integrates file allocation and energy management to maintain least CPU and memory consumption as shown in Figure \ref{Fig:cpu-ram}.

% \begin{figure}[t]
% \centering
% \includegraphics[width=6.5cm]{graphs/cpu.pdf}
% \caption{CPU usage (\%) in different frameworks}
% \label{Fig:cpu}
% % \end{figure}

% % \begin{figure}[ht]
% \centering
% \includegraphics[width=6.5cm]{graphs/ram.pdf}
% \caption{RAM usage (\%) in different frameworks}
% \label{Fig:ram}
% \end{figure}%

\section{Conclusions and Future Work}
\label{S:Conclusion} 

APEX can successfully simulate and provide optimum values of coefficients in the Priority Factor (PF) which can rank the blocks to optimize recovery performance. The model converges to give PF as given in \textit{Section \ref{S:model}}. Using this factor (dynamically optimized by Q-Learning), if we rank all blocks in the disk then it can provide a file allocation system that has a higher data recovery ratio. Thus, at any point if we want to recover a file, it would be recovered to the maximum extent in this file system design compared to existing implementations. The current model gives a recovery performance improvement of 280\%, where performance is determined as mentioned in \textit{Section \ref{S:model}}. Experiments in \textit{Section \ref{S:rr}} show that the APEX  improves the recovery ratio by 678\%, and 32\% compared to the base Ext4 file system and ExtSFR (best among the prior work). Hence, APEX can improve the data recovery with minimum read-write or compute overhead and hence is most apt for resource-constrained edge devices vulnerable to attacks using data stealing malware.

The proposed work only aims to enhance recover-ability of deleted files. This can be extended to secure file-systems from malicious transfer and corruption of files and to cover other types of file-systems like Distributed file-systems. Further, the model can optimize wear levelling in tandem with recovery performance, especially for Flash and SSDs. It can also be extended for emerging Non-Volatile Memories like RRAMs (Resistive Random Access Memory) \cite{RRAM-VAC}, for upcoming energy efficient edge devices. The codes developed and used for data recovery and the Q-Learning Model with simulation results can be found at: \url{https://github.com/HS-Optimization-with-AI}.

\vspace{12pt}

\end{document}